\documentclass[preprint]{aastex}
\usepackage{cite}
\usepackage{graphicx}
\usepackage{amssymb}
\usepackage{amsmath}
\shorttitle{Clustering of the first galaxies}
\shortauthors{Stiavelli \& Trenti}

\begin{document}

\title{The clustering properties of the first galaxies}

\author{Massimo Stiavelli}
\affil{Space Telescope Science Institute, 3700 San Martin Dr., Baltimore MD 21218, USA}

\and

\author{Michele Trenti}
\affil{University of Colorado, Center for Astrophysics and Space Astronomy, 389-UCB, Boulder, CO 80309, USA}

\begin{abstract}
  We study the clustering properties of the first galaxies formed
  in the Universe. We find that, due to chemical enrichment of the inter-stellar
  medium by
  isolated Population III stars formed in mini-halos at redshift
  $z \gtrsim 30$, the (chronologically) first galaxies are composed of
  metal-poor Population II stars and are highly clustered  on small scales. In contrast, chemically
  pristine galaxies in halos with mass $M\sim 10^8 \mathrm{M_{\sun}}$
  may form  at $z<20$ in relatively underdense regions of the
  Universe. This occurs once self-enrichment by Population III in mini-halos is
  quenched by the build-up of an $H_2$ photo-dissociating radiative
  background in the Lyman-Werner bands. We find that these chemically
  pristine galaxies are spatially uncorrelated. Thus,   we expect that
  deep fields with the James Webb Space Telescope may detect clusters of chemically enriched galaxies but individual chemically pristine objects.  We predict
  that  metal-free galaxies at $10 \lesssim z \lesssim 15$ have surface
  densities of about 80 per square arcmin and per unit redshift but
  most of them will be too faint even for JWST. However, the predicted
  density makes these objects interesting targets for searches behind
  lensing clusters.
\end{abstract}


\section{Introduction}

The concept of first galaxies is not as easy to define as that of
first stars and it's partly a matter of convention. A first, somewhat
arbitrary, decision, is the mass scale where an object becomes
a galaxy. A common convention in the context of the first galaxies is
describing galaxies as those objects forming within dark halos
massive enough that their gaseous content can cool by atomic hydrogen (Ly$\alpha$ cooling), i.e.
characterized by a virial temperature greater than $10^4$ K
\citep{ricotti08,WiseAbel08,Greifetal08}. The advantage of this
definition is that it is based on a physical process and makes galaxy
formation relatively independent of the presence of a Lyman-Werner
(LW) background that could hamper the formation of Population
III stars in mini-halos by photo-dissociating molecular hydrogen
\citep{lepp83,tegmark97,haiman97,machacek03,ts09,trenti09b}.

Once we have defined ``galaxy'' we need to define ``first''.  The usual
assumption is that ``first'' refers to a chronological sequence of
events, so that the first galaxies are those that formed at the
earliest redshifts. It is sometimes assumed that the chronologically ``first'' galaxies will also be the least chemically evolved ones.
In fact, these objects are most likely to form in highly
clustered areas and they are also very likely to have been polluted by
metals produced by Population III stars forming in mini-halos
\citep{WiseAbel08,Greifetal08}. Thus, these chronologically first galaxies would
not have a primordial chemical composition. However, we can imagine
another class of galaxies that would form after a LW
background is in place and would then not be polluted by stars in
mini-halos \citep{jimenez06,trenti09b,Johnsonetal08}. These objects could conceivably form stars out of
pristine primordial gas and would then be galaxies made of Population
III stars. Assuming that they can form Population III stars in significant numbers, these objects would be chemically less evolved (i.e. ``younger'') despite forming later in a chronological sense.

This paper is devoted to exploring the validity of the two scenarios
described above and to studying the redshift distributions and
clustering properties of these two classes of objects. These
quantities are relevant for planning surveys to study the first
galaxies with the James Webb Space Telescope and other (future)
facilities, such as thirty-meter class telescopes used for imaging
behind lensing clusters.

The structure of the paper is the following. In Sec. 2 we discuss our
model for galaxy formation during the Dark Ages, in Sec. 3 we present
our results on star formation rates and clustering. Sec. 4 concludes
with prospects for future detections and speculates on the possibility
that metal-free galaxies have already been observed behind a lensing
cluster by \citet{starketal07}.

\section{Galaxy formation during the reionization era}

Our method is based on a combination of numerical simulations and
analytical considerations developed by
\citet{ts07a}, \citet{tss08}, \citet{ts09}, \citet{trenti09b}, and \citet{trentishull2010}. Here we briefly summarize the main
properties of our model. 

\begin{figure} 
\plottwo{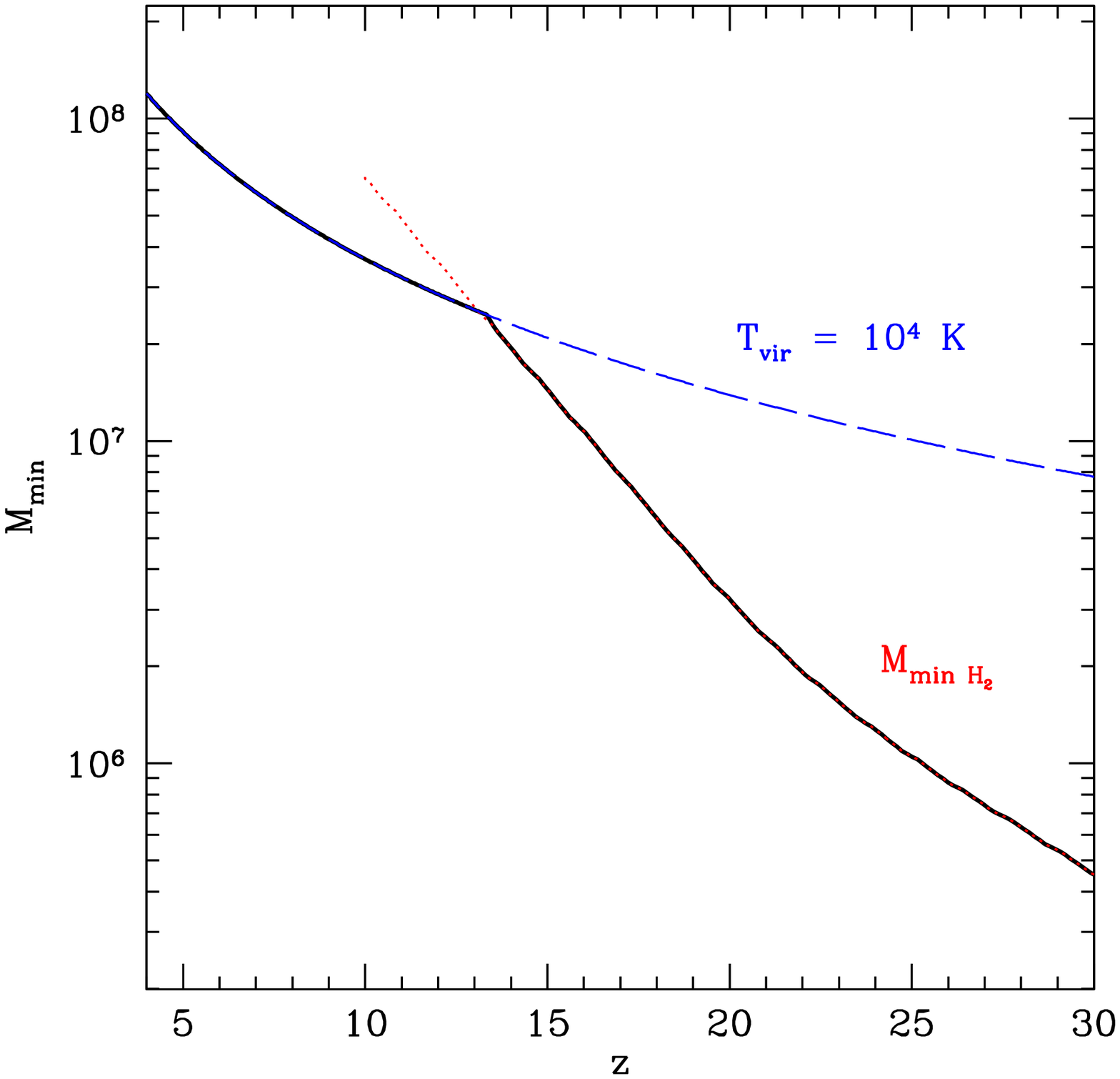}{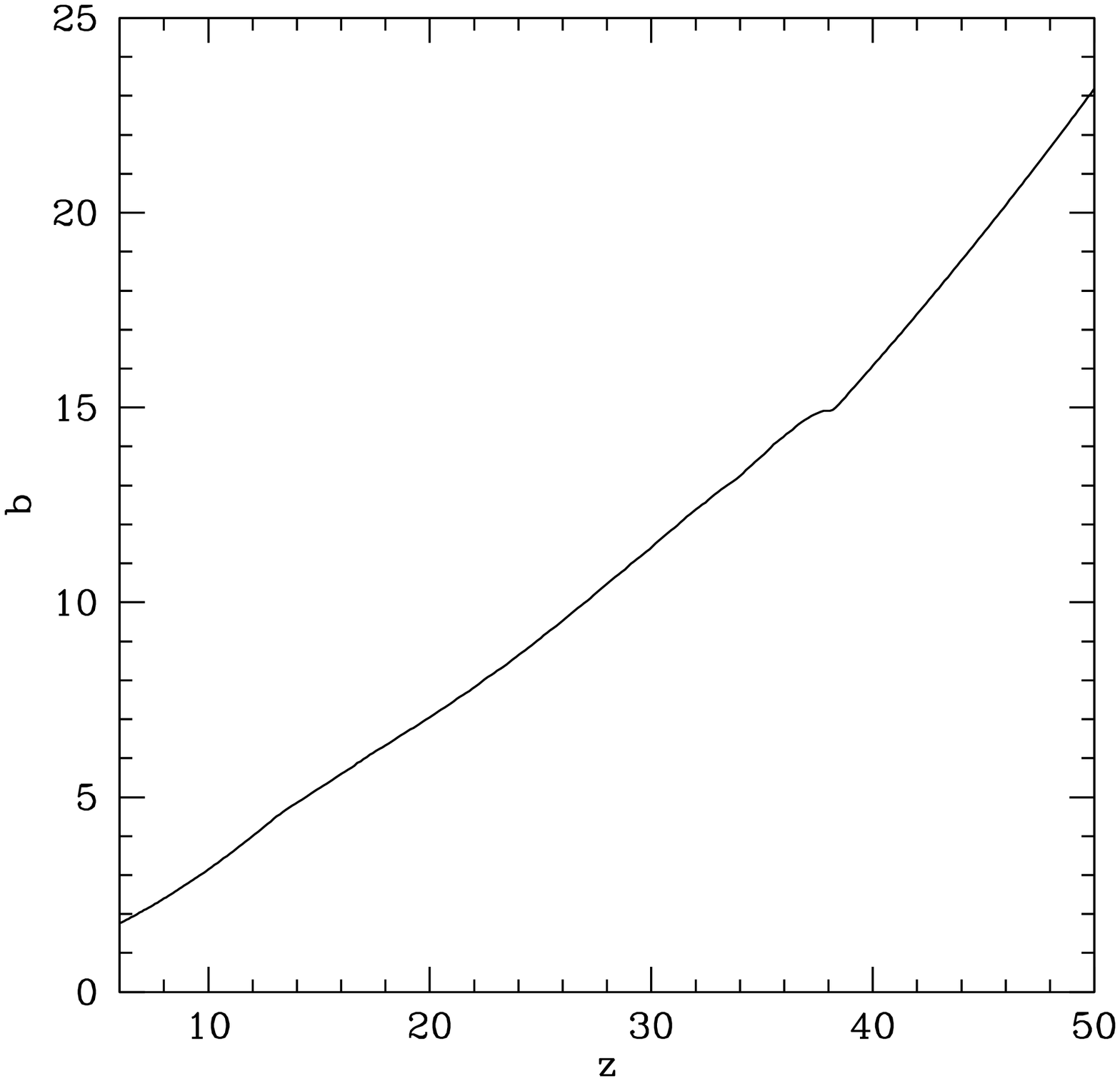}\caption{Left panel. Minimum halo mass
  for star formation (solid black line) required after accounting
  for $H_2$ photo-dissociation due to photons in the Lyman-Werner
  bands (figure from \citealt{trenti09b}). At very high-redshift ($z\gtrsim 30$) the gas can cool in small mini-halos, with $T_{vir} \sim 10^3$ K. As the global star formation rate increases, $H_2$ cooling is suppressed and the minimum mass of a mini-halo capable of forming stars rises steadily. At $z\lesssim 15$ only halos with $T_{vir} \geq 10^4$ K can form stars. Right panel: bias for
  halos at the minimum mass required for star formation as shown in the
  left panel. The bias has been computed using the \citet{st99}
  model.}
\label{fig:bias_st} \end{figure}

We determine the dark matter halo formation rate and merging history
using either extended Press-Schechter modeling based on the
\citet{st99} mass function (see \citealt{ts09}) or high-resolution
cosmological simulations (see \citealt{trenti09b}). We populate
metal-free halos with stars by taking into account the cooling timescales for
atomic and molecular hydrogen, including also radiative feedback due to
LW radiation (\citealt{ts09}, see also \citealt{stiavelli_book}). The
minimum halo mass required for metal-free star formation from our
model is shown as solid black line in left panel of
Fig.~\ref{fig:bias_st}. Metal-free star formation transitions from
mini-halos with virial temperature $T_{vir} \sim 10^3$ K, where
molecular hydrogen cooling is efficient at $z \gtrsim 30$, to more
massive halos ($T_{vir} \gtrsim 10^4$ K) at $z \lesssim 15$, where
Ly-$\alpha$ cooling is possible independently of the radiation
background. The minimum mass we require for star formation is
consistent with the results from detailed hydrodynamic simulations of
metal free star formation in the presence of a radiative LW background
\citep{oshea08}, as discussed in detail in \citet{trenti09b}. If a halo is able to cool and if $T_{\rm
  vir}\sim 10^3$ K \citep{abel02,yoshida03,oshea07}, we assume it forms a
single, massive Pop~III star  drawn from a Salpeter  initial mass function (IMF) in the mass range 50-300 M$_\odot$. When appropriate the star explodes as a Pair Instability
Supernova (PISN) \citep{heger02,scannapieco03}. A constant
fraction of metal-free gas ($f_* \approx 5\times 10^{-3}$) is converted into stars when
Ly$\alpha$ cooling is possible in halos with $T_{\rm vir} \gtrsim
10^4$ K at lower redshift. In this case, we assume a lower
characteristic mass ($O(30) M_{\sun}$) for Population III stars (see \citealt{yoshida06}).

The transition from metal-free to metal enriched star formation
happens at a critical metallicity $Z_{crit}\sim 10^{-3.5} Z_{\sun}$
\citep{bromm04,smith_b09} and might be as low as $Z_{crit} \sim 10^{-6} Z_{\sun}$ in
the presence of dust \citep{schneider06}. A single PISN
provides enough metals to trigger the transition to metal-enriched
star formation in halos with $T_{vir} \sim 10^4$ K (that is
with mass $M \lesssim 3 \times 10^7$ at $z \gtrsim 10$ --- see left
panel of Fig.~\ref{fig:bias_st}). In fact, several studies show that these halos may even reach a metallicity well above $Z_{crit}$ as a result of PISN enrichment \citep{WiseAbel08,karlsson2008,greif2010}.

For metal enriched halos, we assume that a fraction 
$f_* \approx 5\times 10^{-3}$ of gas is converted into stars with a Salpeter IMF in the mass range 1-100 M$_\odot$. In the
analytical model, metal enrichment is based  on a statistical approach making use
of the probability that progenitors of a halo will have formed a
star before the halo collapses, as described in \citet{ts07a}.
In the numerical simulations, we regularly save full snapshots of the system and
use them to construct a detailed halo tree to determine whether a halo is self-enriched.
Through simply counting photons we track the establishment of an
average radiative background, i.e. our models do not include detailed radiative transfer.

The numerical simulations have been run using the Particle Mesh Tree
Code Gadget-2 (Springel 2005). We adopt the fifth year WMAP
concordance cosmology with $\Omega_{\Lambda} = 0.72$, $\Omega_m =
0.28$, $\Omega_b = 0.0462$, $\sigma_8=0.817$, $n_s = 0.96$, $h=0.7$.
We start our main simulation at redshift z=199 using a box with edge
7$h^{-1}$ Mpc, N=1024$^3$ dark matter particles, a mass resolution of
$3.4 \times 10^4$ M$_\odot$ and a force resolution of 0.16 $h^{-1}$
kpc. Halos are identified with a friend-of-friend halo finder (Davis
et al. 1985) using a linking length equal to 0.2 the mean particle
separation. In addition to in situ enrichment by progenitor
mini-halos, in the numerical simulations we also consider enrichment
by winds which is not included in the analytical model. We assume an
outflow speed of $60~\mathrm{km~s^{-1}}$ from halos that contain
metal-enriched galaxies, that is with $T_{vir}\gtrsim 10^4$ K (see
\citealt{trenti09b} for further details). The resulting wind
enrichment is consistent with what is obtained using a
non-cosmological Sedov-Taylor model (e.g. Eq. 8 in
\citealt{tumlinson04}), namely bubble sizes of $\lesssim 150~h^{-1}
\mathrm{kpc}$ at $z\gtrsim 6$. 
Gas polluted by winds in dark matter halos that have not been self-enriched is also likely to reach a metallicity $Z \sim 10^{-3.5} Z_{\sun}$ sufficient to trigger the transition to extremely metal poor (EMP) star formation. In fact, the typical distance traveled by winds before they encounter a non-self enriched halo is of the order of $\sim 50 h^{-1} \mathrm{kpc}$, sufficient to reach the critical level of metal pollution for metal outflows originating from a dwarf galaxy (see Section 2.1 and Figure~3 in \citealt{trenti09b}).

\section{Results}

\subsection{Galaxy Formation Rate}\label{sec:gfr}

In Fig.~\ref{fig:first_gal} we show, as a function of redshift, the rate
of proto-galaxies forming per year and made of chemically enriched
stars (marked as PopII, dashed red line) compared to that of proto-galaxies
made of Population III stars (solid blue line). The figure is derived
from the analytical model for galaxies with virial temperature between
10$^4$ and $2 \times 10^4$ K - roughly corresponding to $\lesssim
10^8$ M$_\odot$ (see left panel of Fig.~\ref{fig:bias_st}) - and
confirms our expectation that chemically enriched galaxies form at
earlier redshifts than galaxies made of population III stars. The
analytical model does not include the effect of enrichment by winds
which will lead to $\sim 30 \%$ lower numbers of pristine objects at
$z \sim 10$ (see Fig.~4 in \citealt{trenti09b}).

In our model, chemically enriched galaxies  peak at a redshift $z\simeq
15$, while galaxies made of Population III stars peak at a redshift $z
\simeq 11$. This trend is in agreement with that shown in
Fig.~1 of \citet{ts09} once one
considers that Pop II formation in that figure was for all halos
regardless of their mass while here it is limited to halos of the
prescribed virial temperature. The presence of a peak at $z\sim 15$ in the enriched
galaxies is due to this mass constraint.

\subsection{Clustering Properties}

In Fig.~\ref{fig:bias_st} we show an analytical estimate of the
bias as a function of redshift for halos that can contain Population
III stars. Clearly the bias decreases rapidly with redshift as the
Population III halos become more common. However, Fig.~\ref{fig:bias_st} does
not distinguish between halos that are polluted by winds and
those that are truly metal free (self-enrichment does not influence
the bias). To address the expected clustering properties of the first
galaxies we need to resort to the numerical simulation results, which
of course carry spatial information on the presence of neighbouring
galaxies capable of polluting pockets of metal free stars. This effect is
important only for redshift $z \lesssim 15$ because at higher redshift
the metal outflows do not have enough time to propagate far from their
host galaxies (see \citealt{trenti09b}).

The cosmological simulation allows us to measure the bias for both the
chemically enriched ``first'' galaxies population and the Population III
galaxies. To quantify their clustering we use the positions of
galaxy-hosting dark matter halos to construct the three dimensional
two-point correlation function $\theta(r)$ in comoving space, defined
as the excess number of pair counts at separation $r$ over those from
a uniform random distribution. We generate a random uniform
distribution of $40000$ points within the simulation box and adopt the
\citet{landy93} estimator:
\begin{equation}
1+\theta(r) = \frac{DD(r) -2DR(r) + RR(r)}{RR(r)}, 
\end{equation}
where $DD(r)$ are the halo-halo pair counts with distance bound in an
annulus centered on $r$. Similarly $DR(r)$ are the halo-random pair
counts and $RR(r)$ the random-random counts within the same annulus. 
 
In Fig.~\ref{fig:tpcf} we show for redshift $z=9.5$ the correlation
function as a function of radius for chemically enriched protogalaxies,
that is for all dark matter halos with $10^4 \mathrm{K}\leq T_{vir}\leq 2\times 10^4$ K that had
past star formation bursts ($\theta_{PopII}$ marked as "Pop II Gal",
solid red) and for galaxies containing Population III stars, that is
dark matter halos with $T_{vir} \geq 10^4$ K that are chemically
pristine with respect to both wind and self-enrichment
($\theta_{PopIII}$, blue line)\footnote{There are no chemically pristine halos with $T_{vir} \geq 2\times 10^4$ K in our simulation because we assume Population III star formation once halos are capable of Ly$\alpha$ cooling \citep{trenti2010}}. We also show the correlation function
for extremely metal poor protogalaxies, at or above the critical metallicity $Z\sim 10^{-3.5}$ Z$_\odot$, enriched by winds but otherwise not
self-enriched, again hosted in halos with  $10^4 \mathrm{K}\leq T_{vir}\leq 2\times 10^4$ K (green
line). In the same figure we
also show the ratio of $\theta(r)_{PopIII}/\theta(r)_{PopII}$ to
quantify the relative bias of these two galaxy populations
\citep{porciani99,st99}. Clearly for radii in excess of a few tens of
kpc the chemically enriched galaxies exhibit stronger clustering than
the galaxies containing Population III stars. The low bias for
Population III galaxies compared to their metal enriched counterparts
extends up to $z \sim 15$ (see Fig.~\ref{fig:tpcf2}).

\section{Discussion and Conclusions}

\begin{figure} 
\plotone{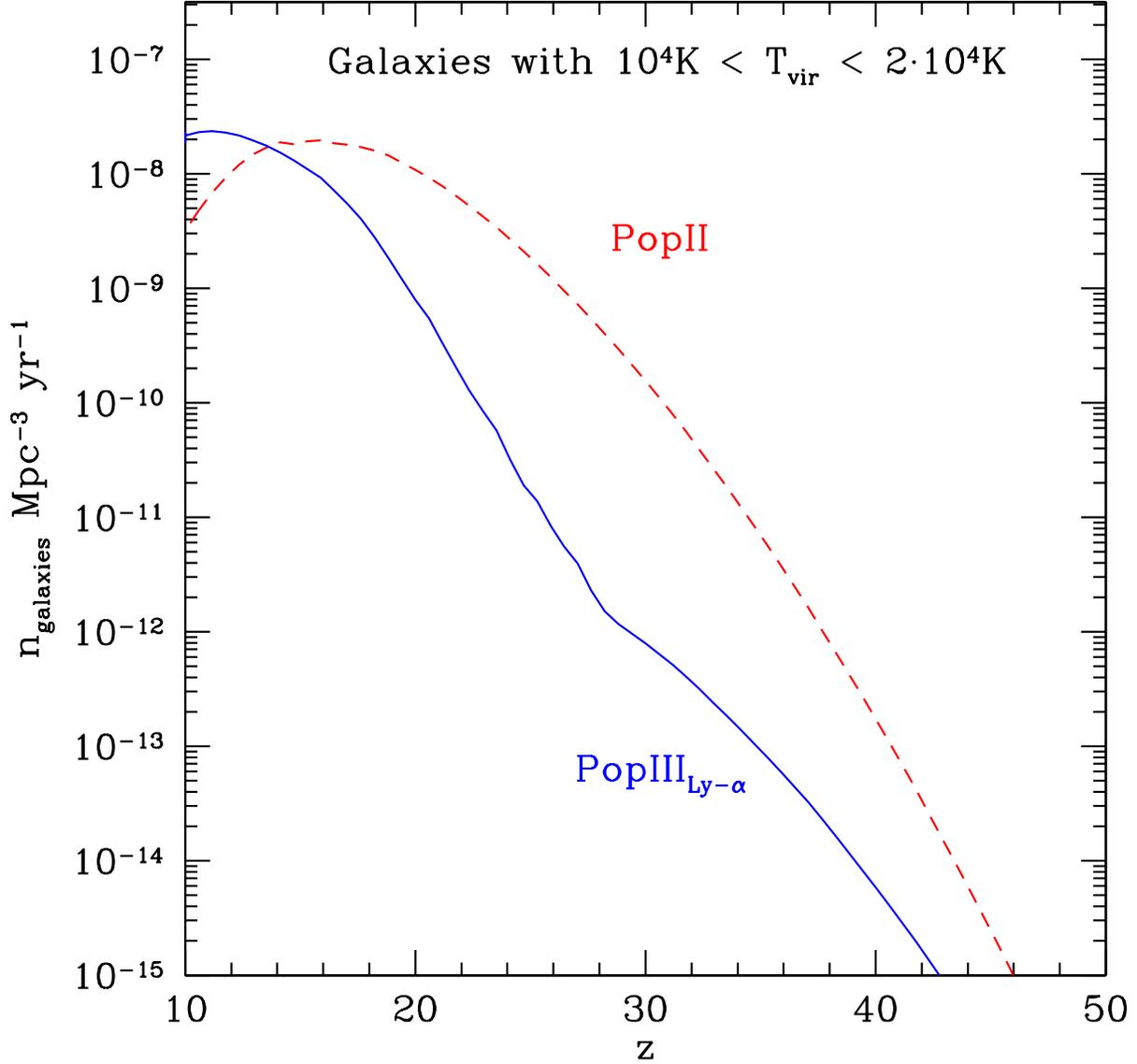}
\caption{Formation rate of
   proto-galaxies per comoving volume (in halos with $10^4~\mathrm{K} \leq T_{vir}< 2
   \cdot 10^4~\mathrm{K}$ that are either chemically pristine (solid blue
 line) or with a low metallicity of the order of the critical one (red dashed line). The
 rate (rest-frame and comoving) has been obtained from an analytical
 Extended Press-Schechter model (see \citealt{ts09}) considering
 self-enrichment only, thus galactic metal outflows are neglected. If
 these were to be considered they would reduce the formation rate of
 PopIII proto-galaxies (see \citealt{trenti09b}). The first galaxies
 to be formed at very high-redshift are very likely to be made of
 Population II (metal enriched) stars.}
\label{fig:first_gal} \end{figure}

\begin{figure} 
  \plottwo{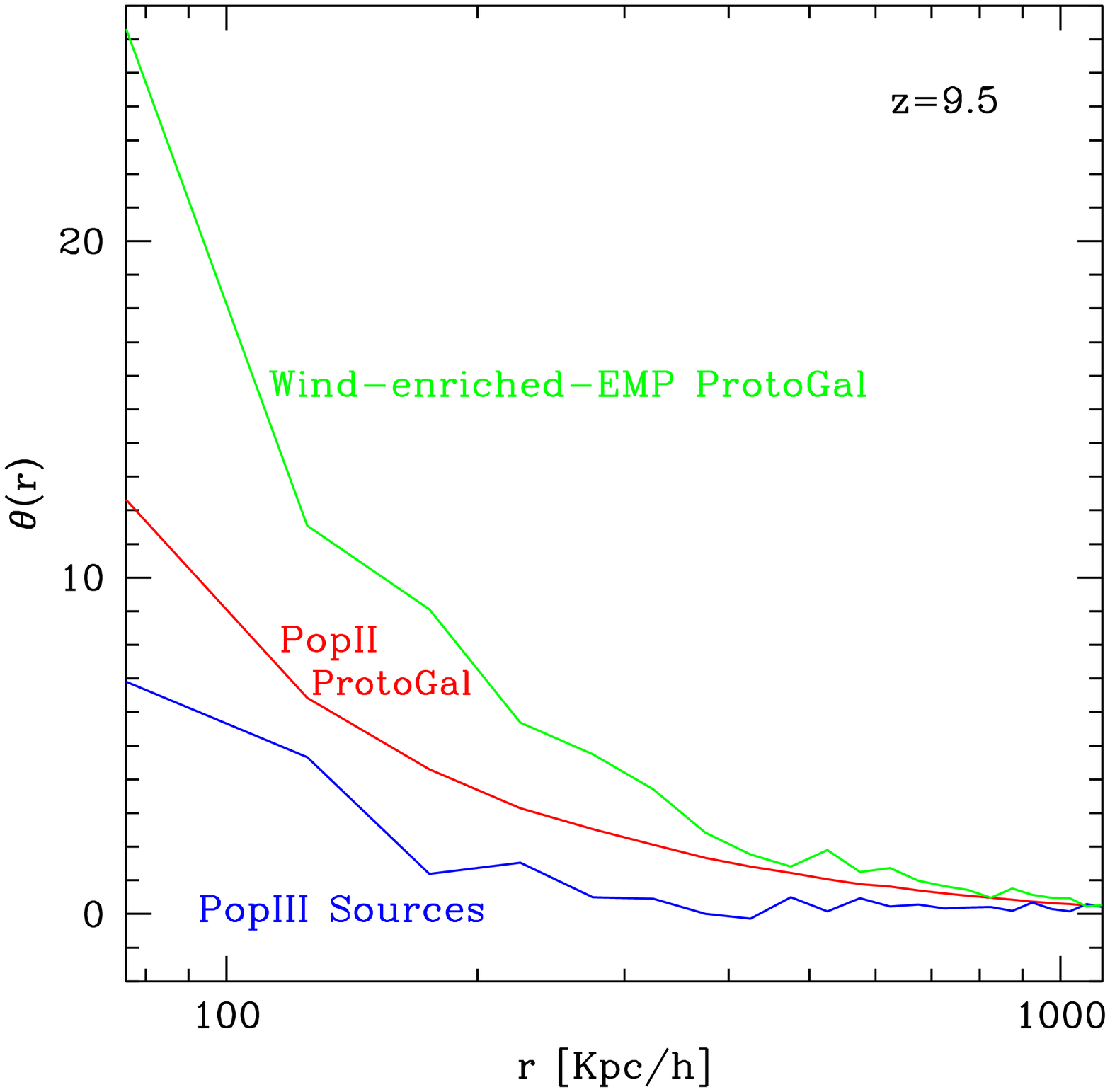}{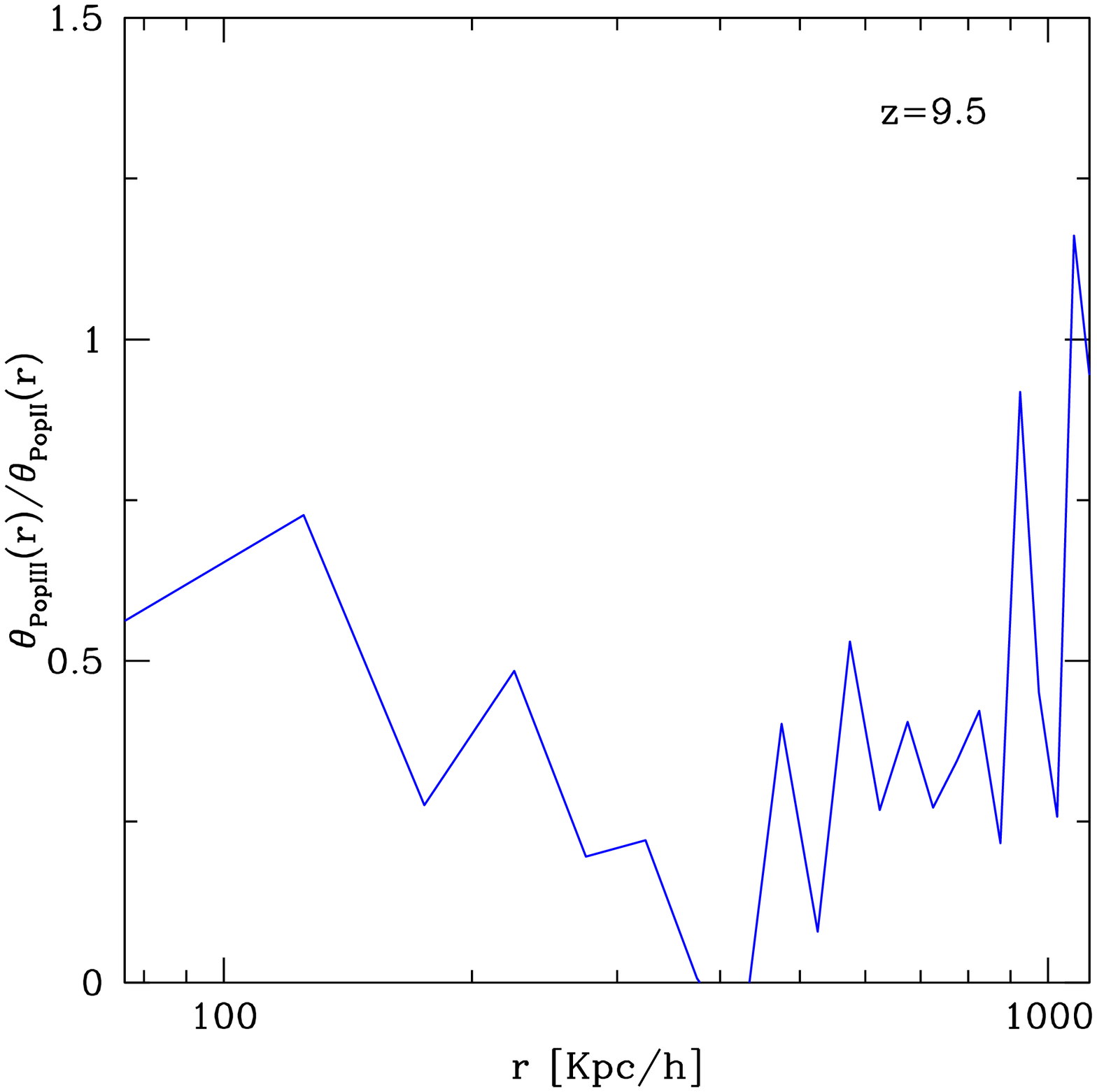}
  \caption{Left
    panel. Three-dimensional two-point correlation function for metal-free (blue) and metal-enriched (red) proto-galaxies, residing in halos with $10^4~\mathrm{K} \leq T_{vir}< 2  \cdot 10^4~\mathrm{K}$, at $z = 9.5$. We also show the correlation for extremely metal poor galaxies (green line) that
    have been enriched by winds. The figure has been obtained from the cosmological
    simulation presented in \citet{trenti09b} and includes metal
    enrichment from outflows propagating at $60~
    \mathrm{km~s^{-1}}$. Lines are an average over 9 simulation snapshots with $9\leq z \leq 10$. 
   Right panel: ratio of the Pop~III to Pop~II correlation
    functions, which further highlights that metal-free proto-galaxies
    are significantly less clustered on small scales ($r\lesssim 1 $ Mpc/h) than their metal-enriched
    counterparts with similar halo mass. }
\label{fig:tpcf} \end{figure}

\begin{figure} 
  \plottwo{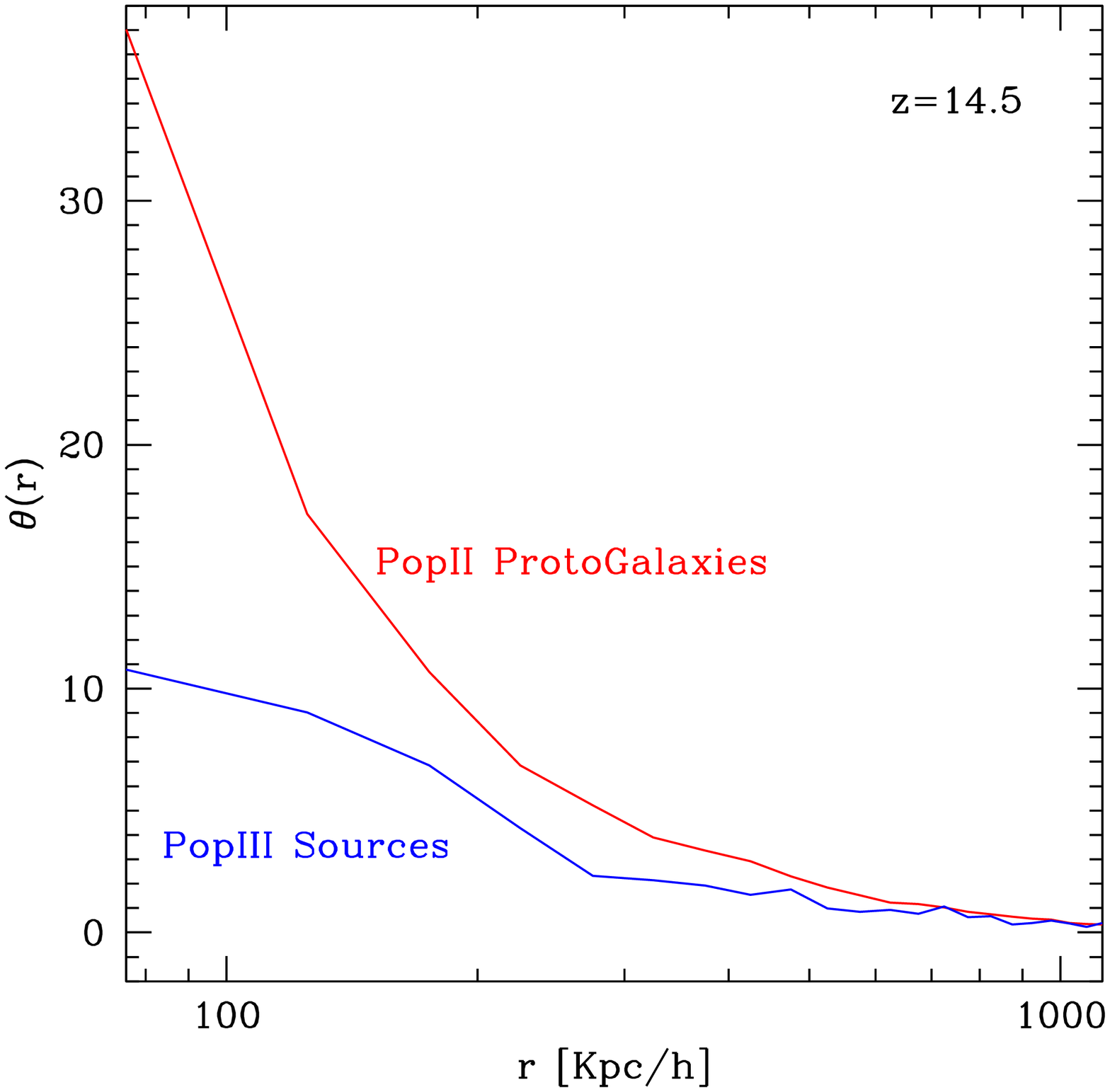}{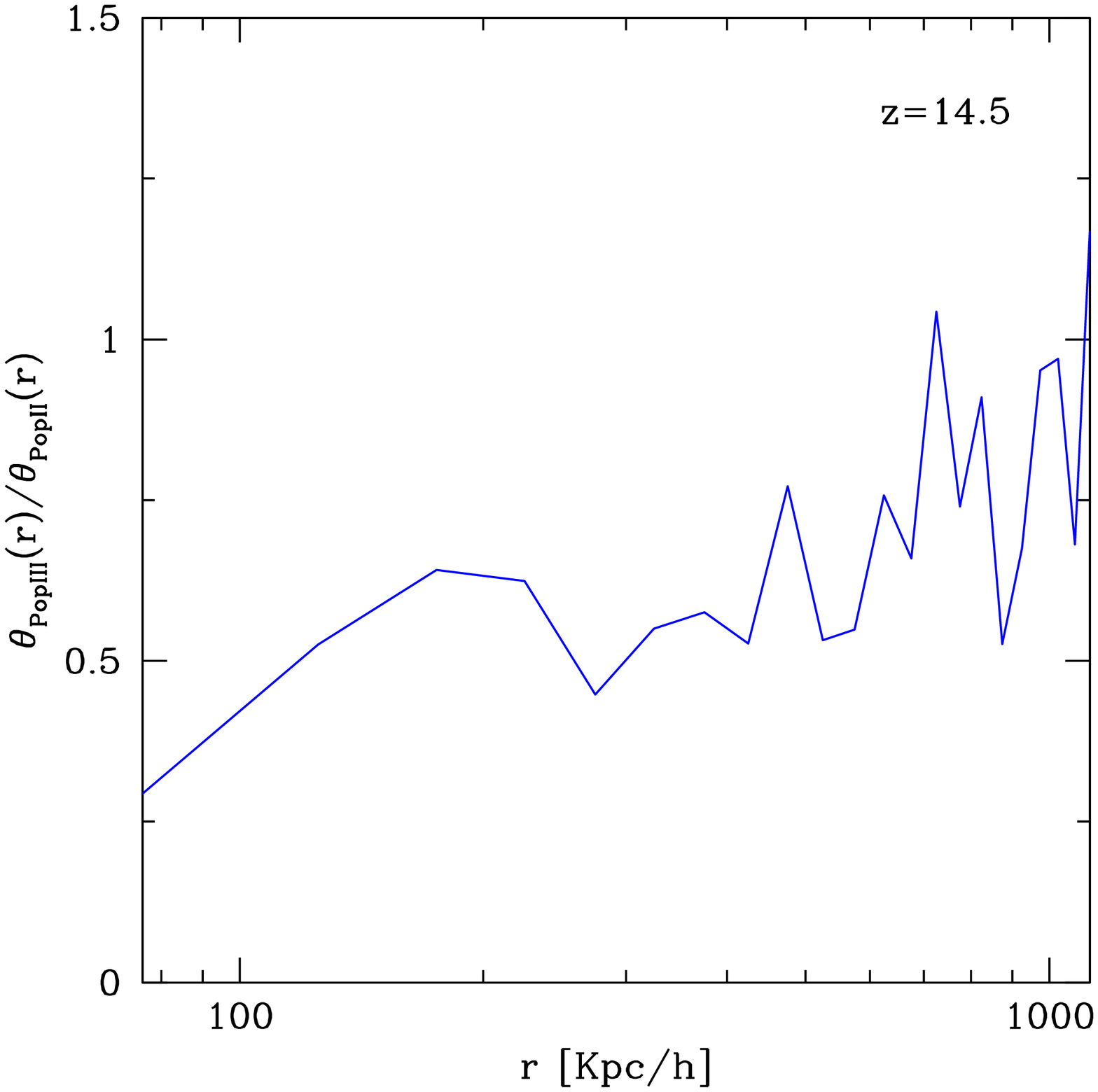}
  \caption{Left panel. Three-dimensional two-point correlation function for metal-free (blue) and metal-enriched (red) proto-galaxies as in Fig.~\ref{fig:tpcf} measured at $z =
    14.5$. The figure has been obtained by averaging over 5 simulation snapshots with  $14\leq z \leq 15$. 
Right panel: ratio of the two-correlation
    functions.}
\label{fig:tpcf2} \end{figure}

Fig.~\ref{fig:tpcf} predicts that galaxies containing Population III
stars will be essentially uncorrelated on scales larger than $\sim150$
comoving kpc. At redshift $z=9.5$, 1 arcsec corresponds to roughly 50
comoving kpc in the transverse direction. This implies that the
typical arcmin scales of the imaging instruments on the James Webb
Space Telescope correspond to a few comoving Mpc and on those scales
we should not detect any significant clustering for galaxies with
primordial chemical composition. In contrast, enriched protogalaxies could be very clustered on small scales, as they live in high density regions. Thus, in an ultradeep field with JWST we expect to detect individual unevolved galaxies 
at $z\simeq10$, but small clusters of chemically evolved objects.  From
Fig.~\ref{fig:first_gal} we see that the rest frame rate of galaxy formation per
comoving volume at $z\simeq10$ is $\sim 2\times 10^{-8}$
Mpc$^{-3}$ yr$^{-1}$ (estimating the halo formation rate from the
simulations gives a similar number). Assuming that the burst of
Population III stars is short lived, we expect that the galaxy will be
visible as Population III galaxy only for a time of the order of the
stellar lifetime (assumed to be $2\times 10^6$ yrs). Considering that
the comoving volume per unit redshift at $z\sim 10$ is $\sim2000$
comoving Mpc$^3$ over an area of one arcmin squared, we predict a
surface density of $~\sim 80$ galaxies per square arcmin  and per unit
redshift. However, there are large uncertainties as to the luminosity
such a galaxy would have. In a deep exposure, the James Webb
Space Telescope should be able to detect a young cluster with a stellar mass
of $10^6$ M$_\odot$\footnote{In contrast, typical Lyman Break Galaxies currently found at $z>7$ are older (stellar age $\gtrsim 100$ Myr \citealt{labbe10}) and reside in dark-matter halos with $M_{halo} \gtrsim 2 \times 10^{10} M_{\sun}$}. Most of the pristine galaxies will have lower
stellar masses. At a virial temperature $T = 10^4$ K at $z\sim 10$ a
galaxy has a halo mass of $\sim3.7 \times 10^7$ M$_\odot$ and only
$\sim 6 \times 10^6$ M$_\odot$ in baryons. It is at this stage unclear
whether the Population III stellar burst in a proto-galaxy would
produce many stars and with what efficiency $\epsilon$, but $\epsilon
\gtrsim 0.15$ is unlikely in the first burst. Even if the majority of
these objects will be too faint, their large number density and the
expected lack of strong correlations are encouraging as they make it
more likely that we could detect some of these objects through a
lensing cluster. A Population III galaxy of $10^6$ M$_\odot$ of stars
formed over a stellar lifetime forms stars at about 0.3 M$_\odot$
yr$^{-1}$ and would have a Ly$\alpha$ luminosity of $\sim3 \times 10^{42}$ erg
s$^{-1}$ \citep{schaerer03} and a Ly$\alpha$ line flux at $z=10$ of
$\sim 1.9 \times 10^{-18}$ erg s$^{-1}$ cm$^{-2}$. These fluxes are
close to what can be detected today in a lensing cluster after modest
amplification and is in fact in agreement with some reports of
detections \citep{starketal07}.

The (chronologically) first galaxies at the same redshift and the extremely poor galaxies are instead
expected to be highly correlated at least on scales of a few
arcsec. Thus, to first order our models predict that one should
observe small clusters of chemically evolved galaxies but only
isolated primordial galaxies.

\acknowledgments

We thank the anonymous referee for comments that have helped improve the paper.
MS acknowledges partial support from NASA JWST IDS grant NAG5-12458.
MT acknowledges partial support from the University of Colorado Astrophysical
Theory Program through grants from NASA (NNX07AG77G) and NSF
(AST07-07474). 


\clearpage

\end{document}